\def\dmthtw{\Delta m^2_{32}}
\def\dmtwon{\Delta m^2_{21}}

\def\eV{{\rm eV}}
\def\ndbd{(\beta \beta)_{0\nu}}
\def\mee{m_{ee}}
\def\misup{m_{\rm IS}^{\rm up}}
\def\mislow{m_{\rm IS}^{\rm low}}

\def\sontw{\sin^2\theta_{12}}

\documentclass[superscriptaddress,aps,nofootinbib,showkeys,showpacs,preprint]{revtex4}
\usepackage{amssymb}
\usepackage{amsmath}
\usepackage{amsfonts}
\usepackage{graphicx}
\usepackage[figuresright]{rotating}
\usepackage{longtable}
\usepackage{bm}
\begin{document}

\title{Neutrino masses from beta decays after KamLAND and WMAP \\ (\small Updated including the NC enhanced SNO data\footnote{Addendum
in page~17 not included in the published version of this paper.
})}

\author{F. R. Joaquim}
\email{filipe@gtae3.ist.utl.pt} \affiliation{Centro de F{\'\i}sica das
Interac\c{c}{\~o}es Fundamentais (CFIF)\\ Departamento de F{\'\i}sica,
Instituto Superior T{\'e}cnico, \\
Av.Rovisco Pais, 1049-001 Lisboa, Portugal}

\begin{abstract}
The first data released by the KamLAND collaboration have
confirmed the strong evidence in favour of the LMA solution of the
solar neutrino problem. Taking into account the ranges for the
oscillation parameters allowed by the global analysis of the
solar, CHOOZ and KamLAND data, we update the limits on the
neutrinoless double beta decay effective neutrino mass parameter
and analyze the impact of all the available data from neutrinoless
double beta decay experiments on the neutrino mass bounds, in view
of the latest WMAP results. For the normal neutrino mass spectrum
the range (0.05-0.23)~eV is obtained for the lightest neutrino
mass if one takes into account the Heidelberg-Moscow evidence for
neutrinoless double beta decay and the cosmological bound. It is
also shown that under the same conditions the mass of the lightest
neutrino may not be bounded from below if the spectrum is of the
inverted type. Finnaly, we discuss how future experiments can
improve the present bounds on the lightest neutrino mass set by
the Troitsk, Mainz and WMAP results.
\end{abstract}
\maketitle

\section{Introduction}
The first results reported by the Kamioka Liquid scintillator
Anti-Neutrino Detector (KamLAND) experiment \cite{Eguchi:2002dm}
have brought new light to the search of the true solution of the
solar neutrino problem. Through large distance measurements, the
KamLAND collaboration has measured, for the first time, reactor
$\overline{\nu}_e$ disappearance within the LMA allowed region for
the neutrino oscillation parameters. Moreover, the
Super-Kamiokande (SK) atmospheric \cite{Fukuda:1998mi} and the KEK
to Kamioka (K2K) accelerator \cite{Ahn:2002up} neutrino
experiments strongly indicate that neutrino flavor oscillations in
the $\nu_\mu \rightarrow \nu_\tau$ channel are the most natural
explanation for $\nu_\mu$ disappearance. These two evidences have
changed our standard picture of particle physics in the sense that
neutrino oscillations require neutrinos to be massive. In addition
to this, neutrinos behave very differently from the other known
elementary particles not only because they are much lighter but
also due to the fact that their mixing pattern differs very much
from the one observed in the quark sector.\\
\indent In spite of all the great achievements of oscillation
experiments, we are still far from a reasonable understanding of
neutrino properties. Among all the unanswered questions in
neutrino physics, the most fundamental one concerns the nature of
neutrinos. In particular, it is of prior importance to know
whether they are Dirac or Majorana particles \cite{Zralek:1997sa}.
While Dirac neutrinos require the existence of highly suppressed
Yukawa couplings, which are difficult to accommodate on
theoretical grounds, small Majorana neutrino masses naturally
arise in the context of minimal extensions of the standard model
where the seesaw mechanism \cite{Yanagida:1979} operates. Still,
our present knowledge of the neutrino sector leaves too much room
for speculation about the neutrino mass generation mechanism.\\
\indent If neutrinos are massive, then the first question which
immediately arises is concerned with the value of their absolute
mass scale. The direct neutrino mass determination method relies
on the detailed analysis of the end-point part of the beta decay
spectrum of some nuclei \cite{Bilenky:2002aw}. At present, the
most stringent experimental bound on the neutrino mass of comes
from the Mainz \cite{Bonn:jw} and Troitsk \cite{Lobashev:uu}
experiments which have set a maximum value for $m_{\nu_e}$ of 2.2
eV. The KATRIN experiment \cite{Osipowicz:2001sq}, which is
planned to start taking data in 2007, will be able to improve the
sensitivity to neutrino masses by approximately one order of
magnitude. \\%
\indent  The search for positive signs in the neutrinoless double
beta [$\ndbd$] decay mode of certain even-even nuclei seems to be,
at present, the most reliable way to look for the Majorana nature
of neutrinos. The phenomenological consequences of $\ndbd$ decays
in the framework of neutrino oscillations have been widely studied
in the literature \cite{Bilenky:2001rz,Petcov:1993rk}. The
extraction of neutrino mass limits through $\ndbd$-decay
measurements involves certain subtleties which are related to the
fact that this method of probing on the absolute neutrino mass
scale strongly depends on the results provided by neutrino
oscillation experiments. Therefore, and in spite of not being
sensitive to it, neutrino oscillation experiments turn out to be
of great importance in the determination of the absolute neutrino
mass scale.

\indent All the outstanding developments in experimental neutrino
physics were accompanied by an equally remarkable evolution of
cosmological experiments. Recently, the results from the Wilkinson
Microwave Anisotropy Probe (WMAP) have brought new insights into
the measurements of a large set of cosmological parameters with an
incredible precision. When combined with the data from the 2
degree Field Galactic Redshift Survey (2dF) \cite{Elgaroy:2002bi},
the WMAP results severely constrain the neutrino masses. In
particular, this simultaneous analysis leads to the following
upper bound on the neutrino contribution to the $\Omega$
cosmological parameter, $\Omega_\nu\,h^2< 0.0076$ where $h$ is the
Hubble constant. This bound, together with the relation
$\sum_im_i=91.5\,\Omega_\nu\,h^2\,(\eV)$ \cite{Kolb:vq} implies
$\sum_i m_i < 0.70~\eV$. Hence, in the framework of three light
neutrino species, one obtains
\begin{align}
\label{boundWMAP} m_i\lesssim 0.23~\eV
\end{align}
for the mass of each neutrino, indicating that the KATRIN
experiment may not have enough sensitivity to measure the electron
neutrino mass. Although this is true for $\beta$-decay experiments
it may not hold for $\ndbd$-decay searches which will be sensitive
to even lower values of neutrino masses, in the future. This
raises the interesting question on how competitive their
results can be when compared with the cosmological ones.\\
\indent Besides the constraints imposed on neutrino masses, the
precise cosmological measurements also place important bounds on
the number of effective allowed neutrino species $N_\nu^{eff}$.
Furthermore, using the range of the baryon-to-photon ratio
$\eta=6.5_{-0.3}^{+0.4}\times 10^{-10}$ and the measurements of
$^4$He primordial abundances, it has been shown  that
$N_\nu^{eff}<3.4$ \cite{Pierce:2003uh}. This result turns out to
be in serious conflict with the evidence for $\overline{\nu}_\mu
\rightarrow \overline{\nu}_e$ oscillations reported by the Liquid
Scintillator Neutrino Detector (LSND) collaboration
\cite{Aguilar:2001ty}. It is well known that the LSND results
require the existence of at least four neutrino species which is
incompatible with the bound $N_\nu^{eff}<3.4$. Moreover, the
neutrino oscillation data indicate that four neutrino oscillation
scenarios are disfavoured when compared with the three neutrino
oscillation scheme \cite{Maltoni:2002ac}. The KARMEN experiment
\cite{Kretschmer:iq} already excludes part of the LSND parameter
region but a conclusive statement about the validity of the LSND
results can only be made by the MiniBooNE experiment
\cite{Stancu:14dr}.

\indent In this paper we update the bounds on the effective
Majorana neutrino mass parameter in view of the latest global
analysis of all the solar, CHOOZ and KamLAND data. This will be
done in the framework of the normal and inverted neutrino mass
schemes. Taking into account the WMAP constraint given in
(\ref{boundWMAP}) and all the presently available neutrino
oscillation and $\ndbd$-decay data we obtain the neutrino mass
bounds for each type of neutrino mass spectrum. Finally, the
impact of future $\ndbd$ decay projects on the determination of
neutrino masses is discussed and the consequences of the first
Heidelberg-Moscow evidence for $\ndbd$ decay, when considered
simultaneously with the WMAP bound on $m_i$, are analyzed.

\section{Neutrino oscillation data: present status}

The presently available neutrino oscillation data, not including
the results from LSND, can be accommodated in the framework of
three mixed massive neutrinos. All the information about neutrino
mixing is enclosed in the leptonic mixing matrix $U$ which relates
neutrino flavor and mass eigenstates in the following way
\begin{align}
\label{flmass} %
\nu_{L\alpha}=\sum_{i=1}^{3}U_{\alpha j}\,
\nu_{Lj}\quad,\quad\alpha=e,\mu,\tau\,,\,j=1,2,3\,.
\end{align}
The mixing matrix $U$ is a $3\times 3$ unitary matrix and its form
depends on whether neutrinos are Dirac or Majorana particles. In
the framework of three light Majorana neutrinos, the matrix $U$
can be parametrized as
\begin{align}
 U  = \left(
\begin{array}{ccc}
c_{12} c_{13} & s_{12} c_{13} & s_{13}  \\
-s_{12} c_{23} - c_{12} s_{23} s_{13} e^{ i \delta} & -c_{12}
c_{23} - s_{12} s_{23} s_{13} e^{ i \delta} &
s_{23} c_{13}e^{i \delta} \\
s_{12} s_{23} - c_{12} c_{23} s_{13} e^{ i \delta} & -c_{12}
s_{23} - s_{12} c_{23} s_{13} e^{ i \delta} & c_{23} c_{13}e^{i
\delta}
\end{array}
\right).\left(
\begin{array}{ccc}
1 & 0 & 0  \\
0 & e^{i\alpha} & 0 \\
0 & 0 &e^{i\beta}
\end{array}
\right)\,,\label{U}
\end{align}
where $s_{ij}\equiv \sin \theta_{ij}$ and $c_{ij}\equiv \cos
\theta_{ij}\,$. The phase $\delta$ is the leptonic Dirac
$CP$-violating phase and $\alpha$ and $\beta$ are Majorana phases
\cite{Bilenky:1980cx}. The Dirac phase $\delta$ can induce
$CP$-violating effects in neutrino oscillations, sizable enough to
be measured by very long baseline neutrino oscillation experiments
in the future \cite{Lindner:2002vt}. On the contrary, oscillation
experiments are blind to the physical effects associated to the
Majorana phases $\alpha$ and $\beta$ . Therefore, the
determination of these phases is only viable in experiments
sensitive to the Majorana nature of neutrinos like $\ndbd$-decay
experiments. Nevertheless, this seems to be a difficult task to
achieve since it requires not only the knowledge of all the
neutrino mass and mixing parameters but also the understanding of
the physics related with $\ndbd$ decays. Namely, the uncertainties
in the nuclear matrix elements involved in the calculation of the
$\ndbd$-decay rates seem to be the major problem on the possible
determination of the Majorana phases \cite{Rodejohann:2002ng}.

\indent The neutrino oscillation experimental results provide us
with information about the neutrino mixing angles $\theta_{ij}$
and mass squared differences $\Delta m_{ij}^2=m_i^2-m_j^2\,$. The
identification of these parameters with the experimentally
measured ones depends on the neutrino mass ordering. In this paper
we will always identify $\theta_{12}$, $\theta_{23}$ and
$\theta_{13}$ with the `solar', `atmospheric' and the CHOOZ
angles, respectively. There are two possible ways of ordering
neutrino masses corresponding to the normal ($m_1<m_2<m_3$) and
inverted ($m_3<m_1<m_2$) spectra. In both cases one can express
two of the neutrino masses as a function of the remaining one and
the $\Delta m_{ij}^2$'s. For the normal neutrino mass spectrum
(NNMS)
\begin{align}
\label{massdef1}m_2=\sqrt{m_1^2+\dmtwon}\,,\,
m_3=\sqrt{m_1^2+\dmtwon+|\dmthtw|}
\end{align}
and for the inverted spectrum (INMS)
\begin{align}
\label{massdef2}m_2=\sqrt{m_3^2+|\dmthtw|}\,,\,
m_1=\sqrt{m_3^2+|\dmthtw|-\dmtwon}\,.
\end{align}
The SK and  K2K neutrino data point towards the existence of
neutrino oscillations in the $\nu_\mu\rightarrow\nu_{\tau}$
channel. The results of these two experiments constrain the
parameters $\theta_{23}$ and $|\dmthtw|$ which, considering the SK
only and SK+K2K data, are found to lie in the ranges
\cite{Fogli:2003th}
\begin{eqnarray} \label{atmdata}%
{\rm SK \;(99\,\%\,C.L.)\;:}& &1.3 \times 10^{-3}\; \text{eV\,}^2
\leq |\dmthtw|  \leq 5.0\times 10^{-3}\;
\text{eV\,}^2\;,\nonumber\\  & &\sin^2 2 \theta_{23}
> 0.85\,,\nonumber\\
{\rm SK+K2K \;(99\,\%\,C.L.)\;:}& &1.4 \times 10^{-3}\;
\text{eV\,}^2 \leq |\dmthtw|  \leq 3.8\times 10^{-3}\;
\text{eV\,}^2\;,\nonumber\\ & &\sin^2 2 \theta_{23}
> 0.85\,,
\end{eqnarray}
with the best-fit values
\begin{eqnarray}
\label{bfatm} & {\rm SK:}&\,|\dmthtw|=2.7 \times 10^{-3}\;
\text{eV\,}^2\,,\,\sin^2 2 \theta_{23}=1\,,\nonumber \\
& {\rm SK+K2K:} &\,|\dmthtw|=2.6 \times 10^{-3}\;
\text{eV\,}^2\,,\,\sin^2 2 \theta_{23}=1\,.
\end{eqnarray}
Although the impact of the K2K data is still not very significant,
the most important fact to retain is that these results are
compatible with the pre-K2K ones.\\
 \indent The presently available data from all solar
and reactor neutrino experiments confirm that the solar neutrino
problem can be explained through the existence of
$\nu_{e}\rightarrow\nu_{\mu,\tau}$ oscillations. After the release
of the first KamLAND results, several global analysis of all the
solar, CHOOZ and KamLAND data have been performed
\cite{Fogli:2002au,Maltoni:2002aw}. As an example we take the
pre-KamLAND and post-KamLAND results obtained in
\cite{Fogli:2002au} which are summarized in Table \ref{table1}.
Besides selecting the LMA region as {\it the} solution to the
solar neutrino problem, the KamLAND experiment significantly
restricts the corresponding oscillation parameter space as can be
seen from Table \ref{table1}. The absence of $\overline{\nu}_e$
disappearance reported by the CHOOZ \cite{Apollonio:1999ae} and
Palo Verde \cite{Boehm:1999gk} experiments imposes severe bounds
on the $\theta_{13}$ angle. The global analysis performed in
\cite{Fogli:2002au} shows that
\begin{align}
\label{chooz} \sin^2 \theta_{13} \lesssim 0.05\quad (99.73\% {\rm
C.L.})\,,
\end{align}
being the best-fit value
\begin{align}
\label{chbest} (\sin^2\theta_{13})_{\rm BF}\lesssim 0.01\,.
\end{align}
\begin{table}
\caption{Allowed ranges for the oscillation parameters $\dmtwon$
and $\sin^2\theta_{12}$ taken from the global analysis of all
solar+CHOOZ and all solar+CHOOZ+KamLAND data performed in
Ref.~\cite{Fogli:2002au}.}
\medskip
\renewcommand{\tabcolsep}{0.6pc}
\begin{tabular}{lcclcc} \hline \hline  \noalign{\flushleft
\hspace{4cm}$2\nu\;{\rm Solar+CHOOZ}$ \hspace{3cm} $2\nu\;{\rm
Solar+CHOOZ+KamLAND}$} \noalign{\medskip} \hline
\noalign{\smallskip} & $\dmtwon~(\times 10^{-4}\,\text{eV}^2)$ &
$\sontw$ &
& $\dmtwon~(\times 10^{-4}\,\text{eV}^2)$ & $\sontw$ \\
\hline 99\% C.L.  &$(\,0.25-4.2\,)$ & $(\,0.21-0.46\,)$ &LMA I &
$(\,0.52-1.0\,)$ & $(\,0.23-0.46\,)$ \\ &  &  &LMA II &
$(\,1.2-2.1\,)$ & $(\,0.23-0.39\,)$\\ \hline 95\% C.L.
&$(\,0.28-2.3\,)$ &$(\,0.24-0.42\,)$
&LMA I & $(\,0.57-0.92\,)$ & $(\,0.24-0.4\,)$\\
  &  &  &LMA II & $(\,1.4-1.8\,)$ &
$(\,0.27-0.33\,)$\\ \hline 90\% C.L.  & $(\,0.3-1.9\,)$ &
$(\,0.24-0.4\,)$ & & $(\,0.6-0.9\,)$ & $(\,0.26-0.4\,)$
\\ \hline Best-fit  & $0.6$ & 0.3
& &0.7 & $0.3$\\
\noalign{\smallskip} \hline \hline \medskip
\end{tabular}
\label{table1}
\end{table}
\indent In spite of being insensitive to the absolute neutrino
mass scale, neutrino oscillation experiments indicate that
neutrinos oscillate with $\dmtwon \ll |\dmthtw|$. This allows the
classification of the neutrino spectrum in hierarchical (HI),
inverted hierarchical (IH) and quasi-degenerate (QD). From
Eqs.~(\ref{massdef1}) and (\ref{massdef2}) one has\footnote{We
will not consider here the cases where $m_1\simeq m_2 < m_3$ and
$m_3 < m_1 \simeq m_2$ which correspond to the partial and partial
inverted mass hierarchy spectra, respectively. The reader is
addressed to Ref.~\cite{Bilenky:2001rz} for a complete analysis on
the subject.}
\begin{align}
\label{HiIh} & {\rm HI:}\;\;m_1 \ll \dmtwon\Rightarrow m_2 \simeq
\sqrt{\dmtwon}\,,\,m_3 \simeq \sqrt{|\dmthtw|}\,,\nonumber \\
& {\rm IH:}\;\;m_3 \ll |\dmthtw| \Rightarrow m_1 \simeq m_2\simeq
\sqrt{|\dmthtw|}\,, \nonumber \\
& {\rm QD:} \;\; m_1 \gg |\dmthtw| \Rightarrow m_1 \simeq m_2
\simeq m_3\,.
\end{align}
Although the presently available neutrino data do not discriminate
between normal and inverted neutrino mass spectra, such a
selection will be possible in future long baseline neutrino
experiments; namely, the detailed study of earth matter effects in
neutrino oscillations will allow for the determination of the sign
of $\dmthtw$ and therefore give us a hint about the mass ordering
of neutrino states \cite{Lipari:1999wy}.

\section{${\bm \ndbd}$ decays and neutrino mass spectra}

\begin{figure*} \caption{Dependence of the effective Majorana neutrino
mass parameter $\mee$ on the lightest neutrino mass for the (a)
NNMS and (b) INMS. The shaded regions were obtained considering
the SK+K2K results and the solar+CHOOZ+KamLAND at 99\% C.L. given
in (\ref{atmdata}) and Table~\ref{table1} respectively, and the
bound (\ref{chooz}) on $s_{13}^2$. The solid lines correspond to
the case where the best-fit values given in (\ref{bfatm}),
(\ref{chbest}) and Table~\ref{table1} are taken into account.}
\begin{tabular}{cc}
\quad \;\;(a) & \quad \;\;(b)\\
\includegraphics[width=8.3cm]{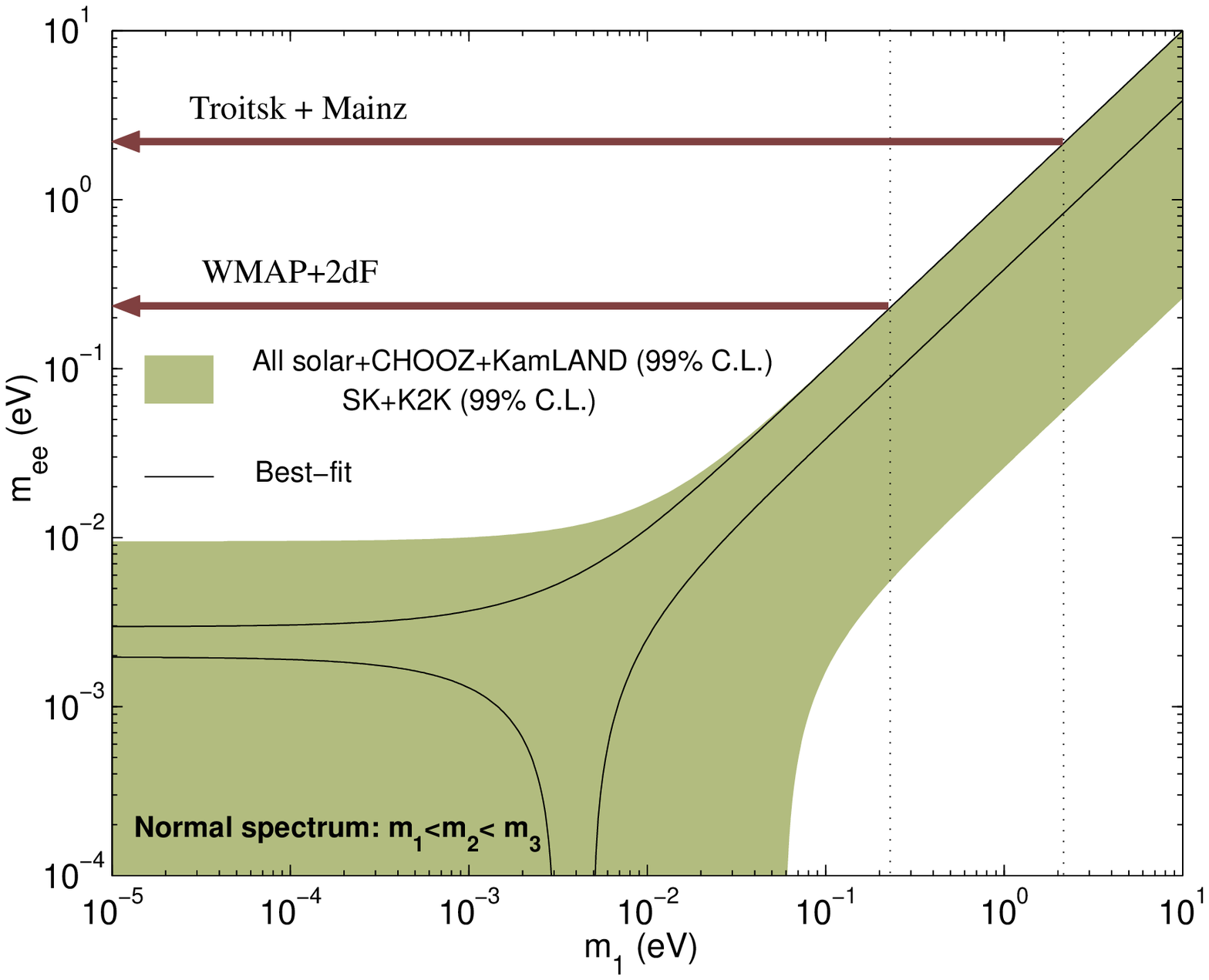}
&\includegraphics[width=8.3cm]{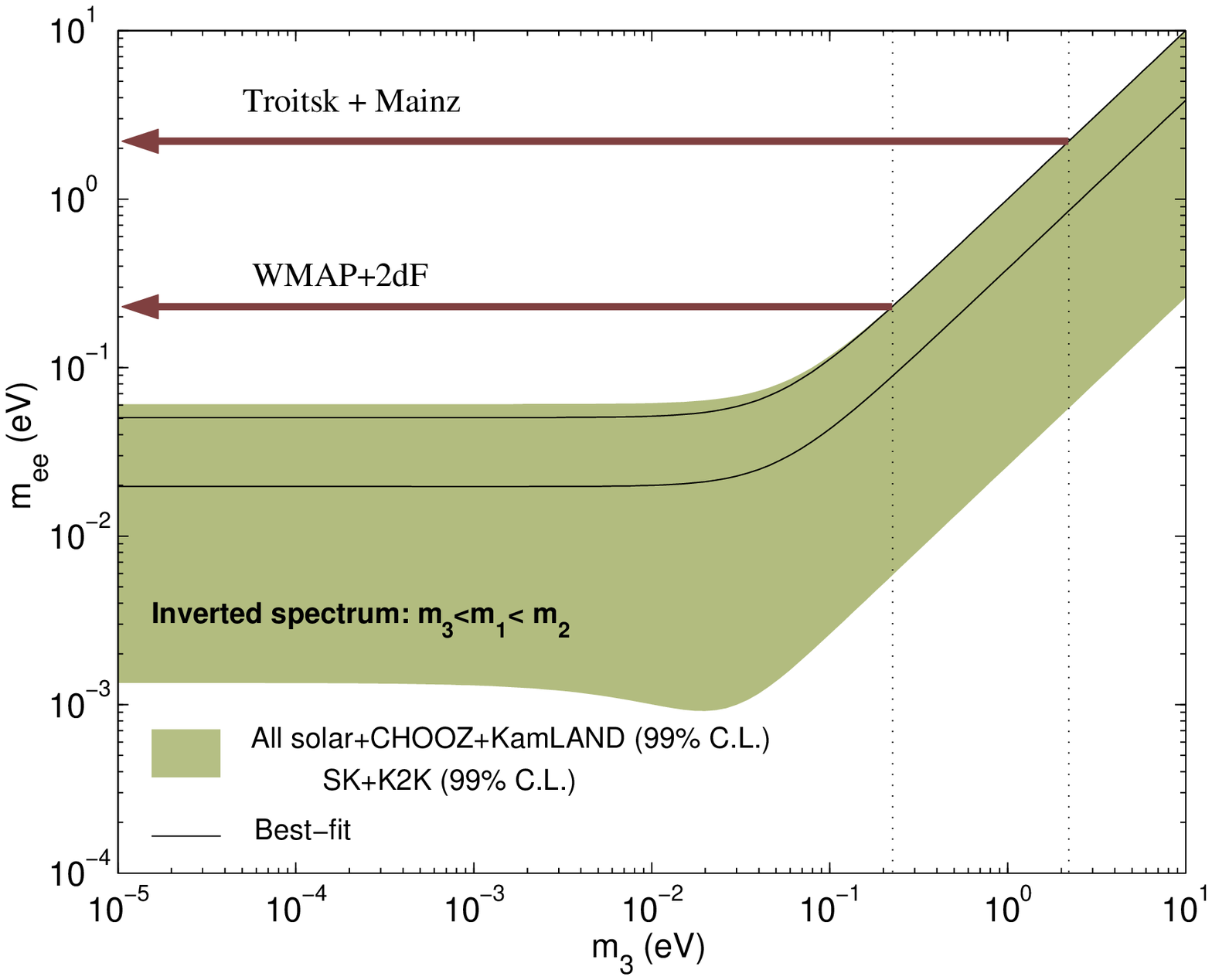}
\end{tabular}\label{fig1}
\end{figure*}

\indent The combined analysis of $\ndbd$-decay and neutrino
oscillation experimental results may be of crucial importance on
the clarification of some aspects related with massive neutrinos.
In particular, the observation of these processes may not only
reveal the Majorana character of neutrinos but also help in the
determination of the absolute neutrino mass scale and spectra. If
these decays occur due to the exchange of virtual massive Majorana
neutrinos, their probability amplitudes are proportional to the
so-called effective Majorana mass parameter
\begin{align}
\label{mee}
\mee=\left|(\mathcal{M}_\nu)_{11}\right|=\left|\sum_{i=1}^{3}
m_iU_{ei}^{\,2}\right|\,,
\end{align}
where $\mathcal{M}_\nu$ is the Majorana neutrino mass matrix,
$m_i$ is the mass of the neutrino mass eigenstate $\nu_i$ and the
$U_{ei}$ are the elements of the first row of the leptonic mixing
matrix. The most significant bounds on the value of the effective
neutrino mass parameter come from the Heidelberg-Moscow
\cite{Klapdor-Kleingrothaus:2000dg} and IGEX \cite{Aalseth:uj}
$^{76}$Ge experiments. Taking into account the uncertainties in
the nuclear matrix elements involved in the determination of the
$\ndbd$-decay amplitudes, one has
\begin{align}
\label{limits1} &\mee \leq
(0.35-1.24)~\eV \quad {\rm (Heidelberg-Moscow)}\,,\\
&\mee \leq (0.33-1.35)~\eV \quad {\rm (IGEX)}.
\end{align}
The reanalysis of the Heidelberg-Moscow data performed in
\cite{Klapdor-Kleingrothaus:2001ke} has been interpreted as an
evidence of $\ndbd$ decay of $^{76}$Ge. The deduced range for
$\mee$ was
\begin{align}
\label{evidence1} \mee=(0.11-0.56)~\eV\quad{\rm 95\%\,C.L.}\,,
\end{align}
which is modified to
\begin{align}
\label{evidence2} \mee=(0.05-0.84)~\eV\quad{\rm 95\%\,C.L.}\,,
\end{align}
if a $\pm 50\%$ uncertainty of the nuclear matrix elements is
considered. The interval
\begin{align}
\label{evidence3} \mee=(0.4-1.3)~\eV\,,
\end{align}
has been obtained in Ref.~\cite{Vogel} using a different set of
nuclear matrix elements. After their publication, the results
presented in Ref.~\cite{Klapdor-Kleingrothaus:2001ke} were
criticized\footnote{For a complete discussion on proofs and
disproofs the reader is addressed to
Ref.~\cite{Klapdor-Kleingrothaus:2003dn} and references therein.}
by some authors \cite{Feruglio:2002af}. At the same time, some
phenomenological implications of this claimed evidence were
explored \cite{Klapdor-Kleingrothaus:2002ip}. In any case, future
experiments will have enough sensitivity to clarify this
situation, and if confirmed, we will surely be in the presence of
the first evidence in favour of the Majorana nature
of massive neutrinos.\\%
 \indent The definition of the effective Majorana neutrino
mass parameter $\mee$ in terms of the physical parameters
$\theta_{ij}$, $m_i$, $\delta$, $\alpha$ and $\beta$, is easily
obtained from Eq.~(\ref{mee}) and the relation
\begin{align}
\label{Mnudef} \mathcal{M}_{\nu}=U^\ast\, \text{diag}\, (m_1 ,
m_2,m_3)\, U^\dagger\,,
\end{align}
which comes from the diagonalization of the $3 \times 3$ Majorana
mass matrix. Parametrizing the mixing matrix $U$ as done in
Eq.~(\ref{U}) one gets
\begin{align}
\label{meedef} \mee=\left|\,m_1U_{e1}^2+m_2 U_{e2}^2 + m_3
U_{e3}^2\,\right|=\left|\,m_1 c_{12}^2 c_{13}^2+m_2
 c_{13}^2 s_{12}^2 e^{-2i\alpha}+m_3 s_{13}^2
 e^{-2i\beta}\,\right|\,.
\end{align}
This expression can be further written in terms of the lightest
neutrino mass $m_1$ ($m_3$) for the NNMS (INMS) and the $\Delta
m^2$'s, taking into account the mass definitions given in
Eqs.~(\ref{massdef1}) and (\ref{massdef2})
\begin{align}
\label{meeNS}\mee^{\rm NS}&=\left|\,m_1 c_{12}^2
c_{13}^2+\sqrt{m_1^2+\dmtwon}\,
 c_{13}^2 s_{12}^2 e^{-2i\alpha}+\sqrt{m_1^2+\dmtwon+|\dmthtw|}
 \,s_{13}^2 e^{-2i\beta}\,\right|\,,
 \end{align}
 \begin{align}
 \label{meeIS}
 \mee^{\rm IS}&=\left|\,\sqrt{m_3^2+|\dmthtw|-\dmtwon}\,c_{12}^2
c_{13}^2+\sqrt{m_3^2+|\dmthtw|}\,
 c_{13}^2 s_{12}^2 e^{-2i\alpha}+m_3s_{13}^2
 e^{-2i\beta}\,\right|\,.
\end{align}
Furthermore, since $\dmtwon \ll |\dmthtw|$, the following
approximations hold
\begin{align}
\label{appmee}
 \mee^{\rm NS}&\simeq m_1\,\left|c_{12}^2
c_{13}^2+\sqrt{1+\frac{\dmtwon}{m_1^2}}\, c_{13}^2 s_{12}^2
e^{-2i\alpha}+\sqrt{1+\frac{|\dmthtw|}{m_1^2}} \,s_{13}^2
e^{-2i\beta}\,\right|\,, \\ \mee^{\rm IS}&\simeq
m_3\,\left|\,\sqrt{1+\frac{|\dmthtw|}{m_3^2}}\,c_{12}^2
c_{13}^2+\sqrt{1+\frac{|\dmthtw|}{m_3^2}}\, c_{13}^2 s_{12}^2
e^{-2i\alpha}+s_{13}^2 e^{-2i\beta}\,\right|\,.
\end{align}
The above equations are valid for any neutrino mass spectrum since
the lightest neutrino mass $m_1$ (or $m_3$) is not constrained.
With the help of Eqs.~(\ref{HiIh}) one can obtain approximate
expressions of the effective neutrino mass parameter $\mee$
depending on the type of neutrino mass spectrum. From
Eqs.~(\ref{meeNS}) and (\ref{meeIS}) and the definitions given in
(\ref{HiIh}) one has
\begin{align}
\label{meespec1} \mee^{\rm HI}&\simeq
\left|\sqrt{\dmtwon}\,s_{12}^2\,c_{13}^2+\sqrt{|\dmthtw|}\,s_{13}^2\,
e^{2i(\alpha-\beta)}\right|\,,
\end{align}
and
\begin{align}
\label{meespec2} \mee^{\rm IH} & \simeq \sqrt{|\dmthtw|}\,c_{13}^2
\left|\,c_{12}^2+s_{12}^2 \, e^{-2i\alpha}\,\right|\,,
 \end{align}
 for the HI and IH neutrino mass spectra, respectively.
 In the case of three quasi-degenerate neutrinos with a common mass
 approximately equal to $m$, Eqs.~(\ref{meeNS}) and (\ref{meeIS})
 reduce to
 \begin{align}
\label{meespec3}
  \mee^{\rm QD} & \simeq m\left|\,c_{13}^2\,(c_{12}^2
+s_{12}^2\,e^{-2i\alpha})+s_{13}^2\,
 e^{-2i\beta}\,\right|\,.
\end{align}
The allowed ranges for the effective neutrino mass parameter in
each case depend not only on the values of the $\Delta m_{ij}^2$'s
and mixing angles but also on the Majorana phases $\alpha$ and
$\beta$. In fact, depending on whether $CP$ is conserved or
violated the value of $m_{ee}$ can drastically change
\cite{Pascoli:2001by}. Here we will only be concerned with the
absolute bounds on $m_{ee}$ which can be easily obtained from
Eqs.~(\ref{meespec1})-(\ref{meespec3}) with the appropriate choice
of the phases $\alpha$ and $\beta$, leading to
\begin{align}
\label{boundHI} (\mee^{\rm HI})_{\rm
low}\simeq\left|\,\sqrt{\dmtwon}\,s_{12}^2\,c_{13}^2-
\sqrt{|\dmthtw|}\,s_{13}^2\,\right| \;,\; (\mee^{\rm HI})_{\rm
up}\simeq  \sqrt{\dmtwon}\,s_{12}^2\,c_{13}^2+
\sqrt{|\dmthtw|}\,s_{13}^2\,,
\end{align}
\begin{align}
\label{boundIH} (\mee^{\rm IH})_{\rm
low}\simeq\sqrt{|\dmthtw|}\,(1-s_{13}^2)(1-2\,s_{12}^2)\;,\;
(\mee^{\rm IH})_{\rm up}\simeq\sqrt{|\dmthtw|}\,(1-s_{13}^2)\,,
\end{align}
\begin{align}
\label{boundQD} (\mee^{\rm QD})_{\rm low}\simeq
m\,\left|\,c_{13}^2\,(1-2\,s_{12}^2)-s_{13}^2\,\right| \; , \;
(\mee^{\rm QD})_{\rm up}\simeq m\,.
\end{align}
\begin{table}
\caption{Allowed ranges for the effective Majorana mass parameter
$\mee$ for each type of neutrino mass spectrum. The values
presented here were determined taking into account the SK+K2K data
summarized in (\ref{atmdata})-(\ref{bfatm}) and the
solar+CHOOZ+KamLAND results given in Table~\ref{table1} and in
(\ref{chooz}) and (\ref{chbest}).}
\medskip
\renewcommand{\tabcolsep}{0.8pc}
\begin{tabular}{llccc}\hline \hline
\noalign{\smallskip}
 & & $\mee^{\rm HI}\,(\times 10^{-3}\,\eV)$
 & $\mee^{\rm IH}\,(\times 10^{-2}\,\eV)$
 & $\mee^{\rm QD}/m$ \\
 \hline
 99\% C.L.
 & LMA I &$\lesssim 7.5$  &$(\,0.3-6.2\,)$
 & $(\,0.026-1\,)$\\
 & LMA II &$\lesssim 8.5$
 & $(\,0.8-6.2\,)$  & $(\,0.16-1\,)$\\
 \hline  95\% C.L. & LMA I
&$\lesssim 6.5$   &$(\,0.7-6.2\,)$ & $(\,0.14-1\,)$\\
 & LMA II &$\lesssim 7.3$   &$(\,1.2-6.2\,)$
  & $(\,0.27-1\,)$\\
 \hline  90\% C.L.  & &
$\lesssim 6.7$  &$(\,0.7-6.2\,)$ & $(\,0.14-1\,)$ \\
 \hline  Best-fit & &
$(\,2.0-3.0\,)$
&$(\,2.0-5.0\,)$ & $(\,0.39-1\,)$\\
\noalign{\smallskip} \hline \hline \medskip
\end{tabular}
\label{table2}
\end{table}
The expression $(\mee^{\rm QD})_{\rm up}$ together with the WMAP
bound on neutrino masses given in (\ref{boundWMAP}) implies $\mee
\lesssim 0.23\,\eV$ which can be interpreted has the
\emph{cosmological} bound on the effective Majorana mass
parameter. Cancellations in $\mee$ can in principle occur in the
HI and QD cases if the following conditions are fulfilled
\begin{align}
\label{cancel} &{\rm HI}\,\, :\;
s_{13}^2=\frac{\sqrt{\dmtwon}\,s_{12}^2}{\sqrt{|\dmthtw|}+\sqrt{\dmtwon}
\,s_{12}^2} \simeq (0.03-0.15)\,,\\
& {\rm QD} : \; s_{13}^2 \simeq \frac{\cos 2\theta_{12}}{1+\cos
2\theta_{12}} \simeq (0.07-0.35)\,.
\end{align}
Comparing these ranges with the bound given in Eq.~(\ref{chooz})
one immediately concludes that cancellations cannot occur if
neutrinos are almost degenerate. Still, the condition $\mee=0$ is
compatible with a hierarchical neutrino mass spectrum. From
Eq.~(\ref{boundIH}) one can see that this can be accomplished in
the IH case if $s_{12}^2=0.5$, which is already excluded by the
data at 99\% C.L.. The allowed ranges for the Majorana mass
parameter are shown in Table~\ref{table2} for each type of
neutrino mass spectrum and the dependence of $\mee$ on the
lightest neutrino mass for the NNMS (INMS) is shown in
Fig.~\ref{fig1}a (b) where the shaded regions indicate the
possible values for $\mee$ in each case. From Fig.~\ref{fig1} and
Table~\ref{table2} it becomes evident that the intervals
determined for $\mee$ may overlap when considering the different
types of neutrino mass spectra. This fact raises the interesting
question on which are the conditions that must be fulfilled in
order to get a clear separation of the different types of neutrino
mass schemes. The imposition of such constraints can be translated
into
\begin{align}
\label{separ1} (\mee^{\rm HI})_{\rm up} < (\mee^{\rm IH})_{\rm
low}\;\;,\;\;(\mee^{\rm IH})_{\rm up} < (\mee^{\rm QD})_{\rm
low}\;\;,\;\; (\mee^{\rm HI})_{\rm up} < (\mee^{\rm QD})_{\rm
low}\,.
\end{align}
Making use of Eqs.~(\ref{boundHI})-(\ref{boundQD}) one can show
that this is equivalent to
\begin{align}
\label{separ} s_{12}^2 < \frac{1-\tan^2
\theta_{13}}{2+\sqrt{\dmtwon/|\dmthtw|}}\;\;,\;\; s_{12}^2 <
\frac{1}{2}\left(1-r_{32}-\tan^2 \theta_{13}\right)\;,\;s_{12}^2 <
\frac{1-\tan^2\theta_{13} \left(1+r_{32}\right)}{2+r_{21}}\,,
\end{align}
respectively, with $r_{32}\equiv \sqrt{|\dmthtw|}/m$ and
$r_{21}\equiv \sqrt{\dmtwon}/m$. Choosing $m \gtrsim 0.2\,\eV$
\cite{Bilenky:2001rz} and $m \lesssim 2.2\,\eV$ (Troitsk+Mainz)
one has, considering the 99\%C.L. bounds on the oscillation
parameters, $\sqrt{\dmtwon/|\dmthtw|} \simeq (0.12-0.39)$, $r_{32}
\simeq (0.02-0.31)$ and $r_{21} \simeq (0.3-7)\times 10^{-2}$.
Together with the inequalities (\ref{separ}) this leads to
\begin{align}
\label{ransepar} s_{12}^2 \lesssim (0.39-0.47) \;\;,\;\; s_{12}^2
\lesssim (0.32-0.49)\;\;,\;\; s_{12}^2 \lesssim (0.45-0.50)\,.
\end{align}
Taking into account that the WMAP result implies $m \simeq
0.23\,\eV$, the first and third relations above remain practically
unchanged whereas the second one is modified to $s_{12}^2 \lesssim
(0.34-0.39)$. This discussion shows that discriminating between
the HI and IH or HI and QD leads to less restrictive conditions
than the one needed for the discrimination between the IH and QD
neutrino mass spectra. The possibility of determining the type of
neutrino mass spectrum has been studied in detail in
Ref.~\cite{Pascoli:2002ae} where the uncertainties in the measured
value of $\mee$ due to the imprecise knowledge of the nuclear matrix
elements were taken into account. In particular, it has been shown that
depending on the uncertainty factor, the above ranges of $s_{12}^2$ get
modified, being even more restrictive.\\
\indent At this point one may wonder if the results in favor of
$\ndbd$-decay reported by the Heidelberg-Moscow collaboration may
help us in the determination of the neutrino mass spectrum type.
Comparing the results plotted in Fig.~\ref{fig1} with the ranges
(\ref{evidence1})-(\ref{evidence3}) it becomes obvious that the HI
spectrum is ruled out. Requiring $(\mee^{\rm IH})_{\rm up}
\lesssim \mee^{\rm min}\simeq 0.05$, one can show that the IH
spectrum is excluded if
\begin{align}
\label{IHevid} |\dmthtw| \lesssim \left[\frac{(\mee)_{\rm
min}}{1-s_{13}^2}\right]^2 \simeq 2.8 \times 10^{-3}\,\eV^2\,.
\end{align}
Hence, considering the SK+K2K ranges for $|\dmthtw|$ given in
(\ref{bfatm}), one concludes that there is still a small window
allowed for the inverted hierarchical neutrino mass spectrum. On
the other hand, the bounds presented in (\ref{evidence1}) and
(\ref{evidence3}) select the QD spectrum as the only possible
scenario for neutrino masses since in these cases $(\mee^{\rm
IH})_{\rm up} < \mee^{\rm min}$.

\section{Neutrino mass bounds from ${\bm \ndbd}$ decays}

In this section we analyze the impact of all the available data
from $\ndbd$-decay experiments on the neutrino mass spectrum,
taking into account the present neutrino
oscillation results and the cosmological bound on neutrino masses.\\
\indent Let us suppose that $\mee$ is found in the range
$\mee^{\rm min} \leq \mee \leq \mee^{\rm max}$. This, together
with Eq.~(\ref{meeNS}) leads to
\begin{align}
\label{condupNS} \mee^{\rm max} \geq \mee^{\rm NS} \geq
c_{13}^2\left|m_1\,(1-2s_{12}^2)-s_{13}^2\sqrt{m_1^2+|\dmthtw|}\,\right|\,,
\end{align}
and
\begin{align}
\label{condlowNS} \mee^{\rm min} \leq \mee^{\rm NS} \leq
m_1+\left(\sqrt{m_1^2+|\dmthtw|}-m_1\right)s_{13}^2\,,
\end{align}
for the NNMS and $\mee^{\rm max} \gg \sqrt{\dmtwon}\,s_{12}^2$.
The above inequalities allow one to find approximate expressions
for the upper and lower bounds of the lightest neutrino mass
$m_1$, which we will denote by $m_{\rm NS}^{\rm up}$ and $m_{\rm
IS}^{\rm low}$, respectively. We obtain in this case
\begin{align}
\label{mupNS} m_{\rm NS}^{\rm up} \simeq \frac{\mee^{\rm max} \,
\cos 2 \theta_{12}\, c_{13}^2+s_{13}^2 \sqrt{(\mee^{\rm
max})^2+|\dmthtw|\,(\cos^2 2 \theta_{12}\, c_{13}^4-
s_{13}^4)}}{\cos^2 2 \theta_{12}\, c_{13}^4-s_{13}^4}\,,
\end{align}
\begin{align}
\label{mlowNS} m_{\rm NS}^{\rm low} \simeq \frac{\mee^{\rm
min}\,c_{13}^2-s_{13}^2\,\sqrt{(\mee^{\rm
min})^2+|\dmthtw|\,(1-2\,s_{13}^2)}}{1-2s_{13}^2}\,.
\end{align}
On the other hand, if  $\mee^{\rm min} \lesssim
\sqrt{\dmtwon}\,s_{12}^2$, the lower bound on $m_1$ does not
exist. This can be understood taking into account that this
condition is compatible with $m_1=0$. In particular, from
Eq.~(\ref{boundHI}) one has $(\mee^{\rm HI})_{\rm low} \lesssim
\sqrt{\dmtwon} s_{12}^2$, which means that for $\mee^{\rm min}
\lesssim \sqrt{\dmtwon} s_{12}^2 \simeq 0.0025\,\eV$ the value of
$m_1$ is not bounded from below. The expressions for $m_{\rm
NS}^{\rm up}$ and $m_{\rm NS}^{\rm low}$ can be further simplified
if the terms proportional to $|\dmthtw|$ are negligible or more
specifically if
\begin{align}
\label{cond} (\mee^{\rm max})^2 \gg |\dmthtw|\,(\cos^2 2
\theta_{12}\, c_{13}^4- s_{13}^4)\,\, , \,\, (\mee^{\rm min})^2
\gg |\dmthtw|\,(1-2\,s_{13}^2)\,.
\end{align}
If this is the case, one gets
\begin{align}
\label{mulNSap} m_{\rm NS}^{\rm up} \simeq \frac{\mee^{\rm max}
}{\cos 2 \theta_{12}\, c_{13}^2-s_{13}^2}
\end{align}
and
\begin{align}
\label{mulNSap1} m_{\rm NS}^{\rm low} \simeq \mee^{\rm min}\,,
\end{align}
which show that $m_{\rm NS}^{\rm up}$ strongly depends on the
$\theta_{12}$ angle while $m_{\rm NS}^{\rm low}$ is only affected
by the value of $\mee^{\rm min}$. Equivalent expressions can be
found for the INMS considering that
\begin{align}
\label{condupIS} \mee^{\rm max} \geq \mee^{\rm IS} \geq
\left|c_{13}^2(1-2s_{12}^2)\sqrt{m_3^2+|\dmthtw|}-m_3\,s_{13}^2\right|\,,
\end{align}
\begin{align}
\label{condlowIS} \mee^{\rm min} \leq \mee^{\rm IS} \leq
\sqrt{m_3^2+|\dmthtw|}\,,
\end{align}
for $\mee^{\rm max} \gg \sqrt{\dmtwon}\,s_{12}^2$. These
inequalities imply
\begin{align}
\label{mupIS} \misup \simeq \frac{\mee^{\rm max} \, s_{13}^2+\cos
2 \theta_{12}\,c_{13}^2 \sqrt{(\mee^{\rm
max})^2-|\dmthtw|(\,\cos^2 2 \theta_{12}\, c_{13}^4- \,
s_{13}^4)}}{\cos^2 2 \theta_{12}\, c_{13}^4-s_{13}^4}\,,
\end{align}
\begin{align}
\label{mlowIS}\mislow \simeq \sqrt{(\mee^{\rm
min})^2-|\dmthtw|}\,,
\end{align}
which reduce to the expressions given in Eq.~(\ref{mulNSap}) in
the limit where the terms proportional to $|\dmthtw|$ can be
neglected. It can easily be seen from the above expression that
for $\mee^{\rm min} \lesssim \sqrt{|\dmthtw|_{\rm max}}\simeq
0.06\,\eV$ there is no lower bound on $m_3$ in the INMS case.\\
\indent Let us now discuss the implications of the
Heidelberg-Moscow experimental result given in (\ref{limits1}) on
the values of neutrino masses.
\begin{figure*}
\caption{(a) Allowed region for the upper bound on the lightest
neutrino mass for both the NNMS and INMS  when the
Heidelberg-Moscow result $\mee \lesssim (0.35-1.24)\,\eV$ is taken
into account. The horizontal dash-dotted line indicates the bound
on $m^{\rm up}$ set by the Troitsk and Mainz experiments. The
vertical dashed lines delimit the $s_{12}^2$ ranges given in
Table~\ref{table1}, and the vertical dotted line corresponds to
the best-fit value $(s_{12}^2)_{\rm BF}=0.3$. The limit on $m^{\rm
up}$ imposed by the WMAP+2dF results is also shown. (b) Values of
$(s_{12}^2)_{\rm max}$ under which the bounds $m^{\rm
up}=2.2\,\eV$ (Trotsk+Mainz) and $m^{\rm up}=0.23\,\eV$ (WMAP+2dF)
can be improved for a given $\mee^{\rm max}$. The horizontal
dash-dotted lines delimit the presently allowed regions for
$s_{12}^2$ at 99\% C.L. and the vertical dotted lines correspond
to the sensitivities of future $\ndbd$-decay experiments.}
\begin{tabular}{cc}
\quad \quad(a) & \quad \quad (b)\\
\includegraphics[width=8.3cm]{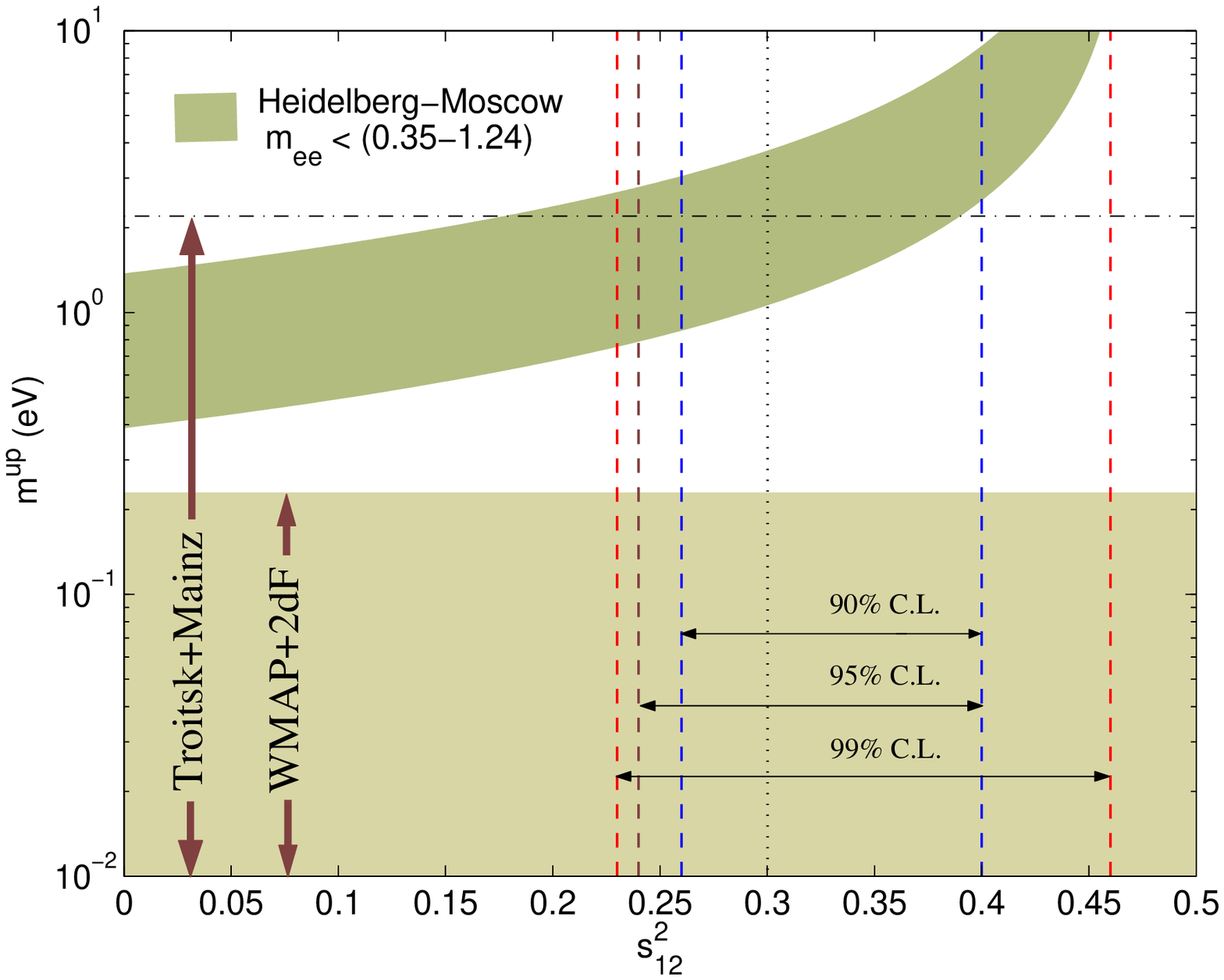}
&\includegraphics[width=8.3cm]{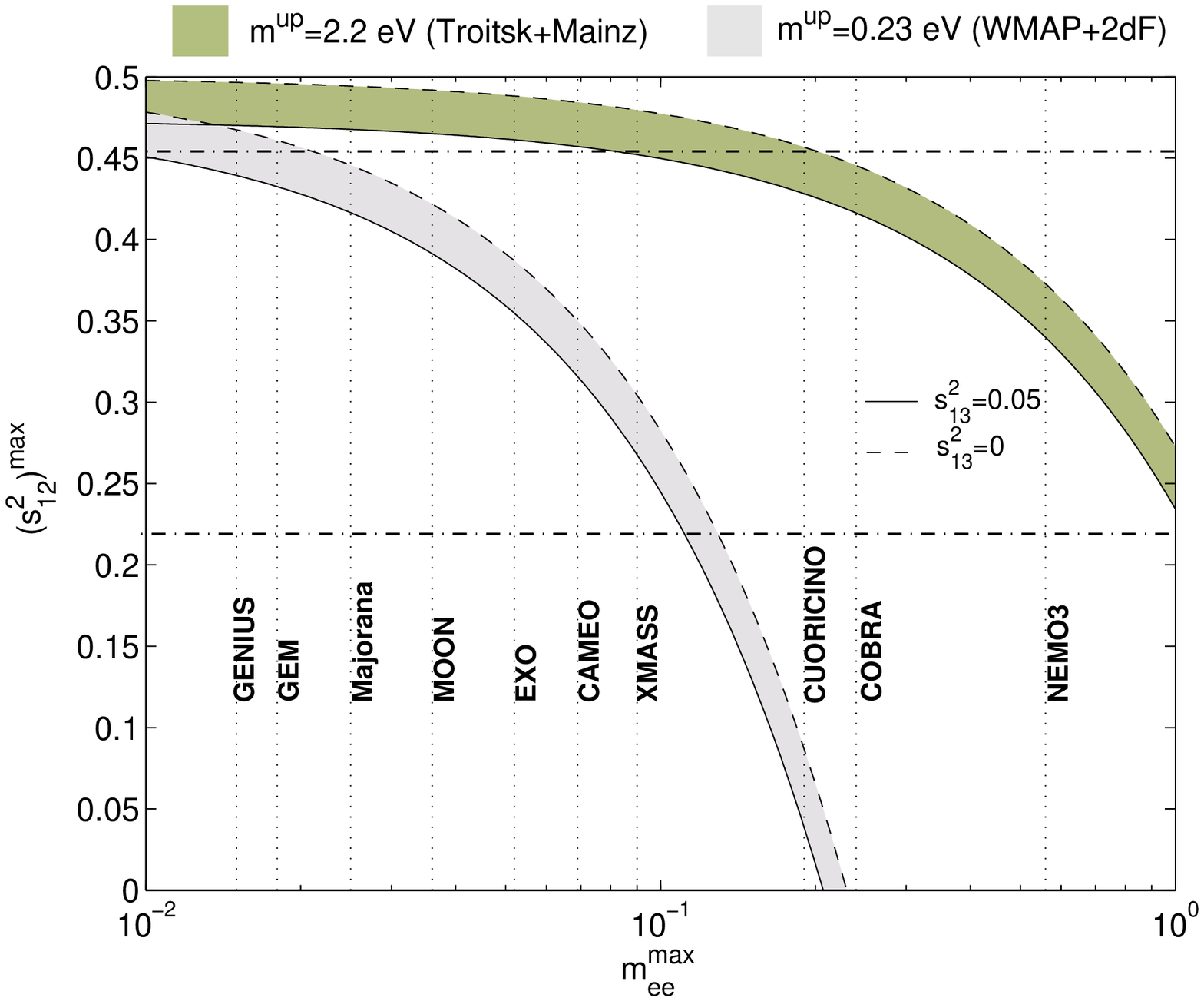}
\end{tabular}\label{fig2}
\end{figure*}
In this case $\mee \lesssim (0.35-1.24)\,\eV$ and therefore only
an upper bound on the neutrino masses can be set. In order to
compute it we can use the first expression in (\ref{mulNSap}) for
both NNMS and INMS since for this range of $\mee^{\rm max}$ the
first inequality in Eq.~(\ref{cond}) is verified. Therefore, we
will denote both $m_{\rm NS}^{\rm up}$ and $m_{\rm IS}^{\rm up}$
just by $m^{\rm up}$. In Fig.~\ref{fig2}a we show the allowed
region for $m^{\rm up}$ as a function of $s_{12}^2$, obtained for
$s_{13}^2=0.05$. Taking into account the Troitsk and Mainz bound
on $m^{\rm up}$ and the 99\% C.L. allowed range for $s_{12}^2$ one
can conclude that $\mee^{\rm max} \lesssim 1\,\eV$. This can be
seen putting $s_{13}^2=0.05$ and $m_{\rm NS}^{\rm up}=2.2\,\eV$ in
Eq.~(\ref{mulNSap}). Using the best-fit value $s_{12}^2=0.3$ and
the bound (\ref{chbest}) one gets $\mee^{\rm max} \lesssim
0.9\,\eV$. Considering $\mee \lesssim 0.35\,\eV$ one can see that
the value of $m^{\rm up}$ extracted from $\ndbd$ decay is smaller
than the Troitsk and Mainz bound for $s_{12}^2 \lesssim 0.4$.
Alternatively, with $s_{12}^2=0.3$ and $s_{13}^2 =0.01$ we get
$m^{\rm up} \simeq 0.9\,\eV$ for $\mee^{\rm max}=0.35\,\eV$. From
Fig.~\ref{fig2}a we can also see that the bounds on $m^{\rm up}$
are always more conservative than the WMAP bound.\\
\indent The sensitivity to $\mee$ is expected to be improved by
several future experiments like the NEMO3 experiment
\cite{Marquet:eq}, which has started to take data last year, and
the CUORICINO project \cite{Fiorini:gj} to be operative this year.
Other experimental setups are being planned or under construction
with the goal of reaching sensitivities of $\mee \simeq 0.01\,\eV
$ \cite{ndbdfuture}. Taking this into consideration, it is
interesting to analyze how these upcoming experiments can improve
the present bounds on $m^{\rm up}$. It can be shown that in order
for this to happen the following condition has to be verified
\begin{align} \label{s12max}
s_{12}^2\lesssim \frac{1}{2}\left(1- \frac{\mee^{\rm
max}+\sqrt{|\dmthtw|+(m^{\rm up})^2}\,s_{13}^2}{m^{\rm
up}\,(1-s_{13}^2)}\right) \simeq \frac{1}{2}\left(1-
\frac{\mee^{\rm max}/m^{\rm up}+s_{13}^2}{(1-s_{13}^2)}\right)
\equiv (s_{12}^2)_{\rm max}\,,
\end{align}
where the last expression corresponds to the limit $m_{\rm up} \gg
\sqrt{|\dmthtw|}$. In Fig~\ref{fig2}b we show the dependence of
$(s_{12}^2)_{\rm max}$ on the values of $\mee^{\rm max}$ for
$m^{\rm up}=2.2\,\eV$ (Troitsk+Mainz) and $m^{\rm up}=0.23\,\eV$
(WMAP+2dF). The vertical dotted lines indicate the sensitivities
of some future
$\ndbd$-decay experiments. \\
\begin{figure*} \caption{Allowed
ranges for the lightest neutrino mass $m_1$ in the NS case, when
the latest results of the Heidelberg-Moscow experiment are
considered. The shaded region corresponds to the interval of
$\mee$ given in (\ref{evidence1}) and the  dash-dotted and solid
lines to the ranges (\ref{evidence2}) and (\ref{evidence3}),
respectively.}
\includegraphics[width=9cm]{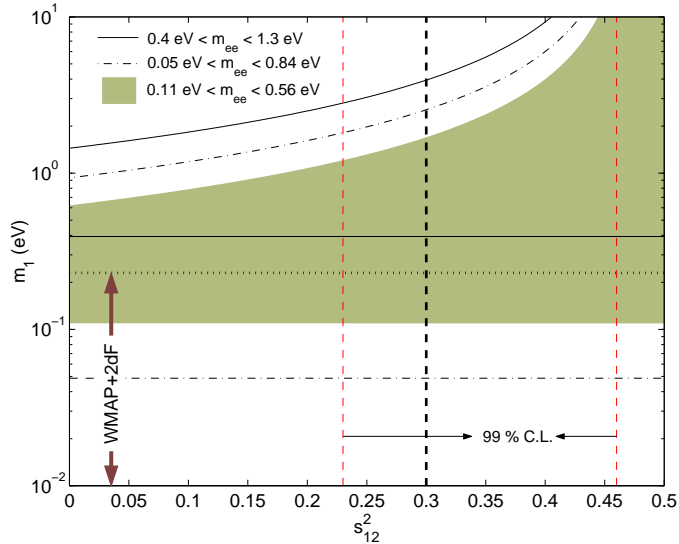}
\label{fig3}
\end{figure*}
\indent Turning now our attention to the evidence for $\ndbd$
decay reported by the Heidelberg-Moscow collaboration and taking
into account the $\mee$ ranges given in
(\ref{evidence1})-(\ref{evidence3}), we can use
Eqs.~(\ref{condupNS}) and (\ref{condlowNS}) to compute the values
of $m_{\rm NS}^{\rm up}$ and $m_{\rm NS}^{\rm low}$ which are
shown in Fig.~\ref{fig3}. The horizontal lines correspond to the
values of $m_{\rm NS}^{\rm low}$ for $\mee^{\rm
min}=0.05,0.11,0.4\,\eV$. We can see that the lower bounds on the
lightest neutrino mass in each case obey the relation $m_{\rm
NS}^{\rm low} \simeq \mee^{\rm min}$, as already shown in
Eq.~(\ref{mulNSap1}). Consequently, the lower bound of $\mee$
given in (\ref{evidence3}) is in conflict with the WMAP result.
Considering the results (\ref{evidence1}) and (\ref{evidence2}),
together with the cosmological bound on neutrino masses, one has
for the NNMS
\begin{align}
\label{m1rangeNS} 0.05\,\eV \lesssim m_1 \lesssim 0.23\,\eV\,.
\end{align}
Regarding the inverted neutrino mass spectrum, we note that the
above results obtained for $m^{\rm up}$ remain valid, as already
discussed before. However, the situation changes for $m_{\rm
IS}^{\rm low}$ in the sense that now the lower bound on the
lightest neutrino mass may not exist. Considering $\mee^{\rm
min}=0.1,\, 0.4\,\eV$ we obtain $m_{\rm IS}^{\rm low} \simeq
0.1,\,0.4\,\eV$ since for these values of $\mee^{\rm min}$ we can
neglect $|\dmthtw|$ in Eq.~({\ref{mlowIS}}). On the other hand,
from Eq.~(\ref{mlowIS}) and the SK+K2K allowed ranges for
$\dmthtw$ given in ({\ref{atmdata}}) one has $m_{\rm IS}^{\rm low}
\lesssim 1.1 \times 10^{-3}\,\eV$ for $\mee^{\rm min}=0.05$.

\section{Conclusions}
In this paper we have focused on the implications of the available
data from $\ndbd$-decay experiments in the light of the latest
neutrino oscillation and WMAP results. We have briefly commented
on the possible occurrence of cancellations in the effective
Majorana neutrino mass parameter taking into account the allowed
ranges for the neutrino oscillation parameters at 99\% C.L. given
in Refs.~\cite{Fogli:2003th} and \cite{Fogli:2002au}. We conclude
that cancellations are only possible for the HI neutrino mass
spectrum. However, this is no longer true if one relies on the
best-fit values given in (\ref{bfatm}), (\ref{chbest}) and
Table~\ref{table1} since in this case the condition $\mee \simeq
0$ cannot be fulfilled for any of the neutrino mass schemes
considered here. As for the extraction of neutrino mass bounds
from the presently available $\ndbd$-decay data we have seen that,
while the establishment of an upper bound on the mass of the
lightest neutrino strongly depends on the value of $s_{12}^2$, the
lower bound is only sensitive to $\mee^{\rm min}$, for the NNMS.
In particular, the range $0.05\,\eV \lesssim m_1 \lesssim
0.23\,\eV$ is obtained if one considers the intervals for $\mee$
given in (\ref{evidence1})-(\ref{evidence3}) together with the
WMAP bound. In the INMS case, the knowledge of $|\dmthtw|$ may be
relevant for the lower bound on the lightest neutrino mass since
in this case the lower bound on $m_3$ may not exist if a $\pm50\%$
uncertainty in the nuclear matrix elements is taken into account.
We have also discussed how the upcoming $\ndbd$-decay experiments
may improve the WMAP result and concluded that, again, the key
point relies on the measurement of $s_{12}^2$ which is expected to
be improved by the future KamLAND and BOREXINO \cite{Shutt:rg}
data. Nevertheless, a sensitivity of $\mee \simeq 0.09\,\eV$ is
required if the best-fit value $s_{12}^2=0.3$ is considered.
Finally, we would like to remark that the cosmological bound on
the absolute neutrino mass scale has important consequences for
the future prospects in the study of $\ndbd$-decay searches. On
the other hand, the bound $m_i < 0.23\,\eV$ is not encouraging for
the KATRIN experiment which will be sensitive to $m_i \gtrsim
0.35\,\eV$. In any case, one should remember that the kind of
analysis presented here is based on the assumption that $\ndbd$
decays are mediated by the exchange of massive Majorana neutrinos.
However, other mechanisms can give rise to these processes
\cite{Vergados:pv}. This possibility opens a new challenging
question, namely how one can identify the physics behind $\ndbd$
decays.

\section{Addendum: Update with the NC enhanced SNO data}

Recently, the SNO collaboration has released the first data from
the salt enhanced NC measurements \cite{Ahmed:2003kj}. Global
analysis of all the solar neutrino and KamLAND data
\cite{Balantekin:2003jm,Fogli:2003vj,Maltoni:2003da,Aliani:2003ns,Creminelli:2001ij,deHolanda:2003nj}
show that, after including the new SNO data, the high-$\Delta m^2$
region ($\Delta m^2 > 10^{-4} \,\eV^2$) of the neutrino
oscillation parameter space is now only accepted at the $3\sigma$
level. Moreover, maximal mixing in the 1-2 neutrino sector is
ruled out at more than $5\sigma$.

In this addendum we update the allowed ranges for $\mee$ obtained
in the published version of this paper, taking into account the
new ranges for the neutrino oscillation parameters (see also
Refs.~\cite{Murayama:2003ci}). We consider the following results
obtained in Ref.~\cite{Maltoni:2003da} at the $2\sigma\,(3\sigma)$
levels
\begin{align}
\label{valleresults}%
&6\,(5.4)\times 10^{-5}\,\eV^2 < \dmtwon < 8.4\,(9.5)\times
10^{-5}\,\eV^2\;\;,\;\; 0.25\,(0.23) < s_{12}^2 <
0.36\,(0.39)\nonumber \\
&1.8\,(3.3)\times 10^{-3}\,\eV^2 < \dmthtw < 1.4\,(3.7)\times
10^{-3}\,\eV^2\,,\nonumber \\
& s_{13}^2 \leq 0.035\,(0.054)\,.
\end{align}
For the best-fit values of the above parameters one has
\cite{Maltoni:2003da}
\begin{align}
\label{bestsnovalle}%
\dmtwon = 6.9\times 10^{-5}\,\eV^2\,\,,\,\, \dmthtw = 2.6\times
10^{-3}\,\eV^2\,\,,\,\, s_{12}^2=0.30\,\,,\,\, s_{13}^2=0.006\,.
\end{align}
\begin{figure*} \caption{The same as in Fig.~\ref{fig1} for the parameter ranges shown in (\ref{valleresults}) and
(\ref{bestsnovalle}).}
\begin{tabular}{cc}
\quad \;\;(a) & \quad \;\;(b)\\
\includegraphics[width=8.3cm]{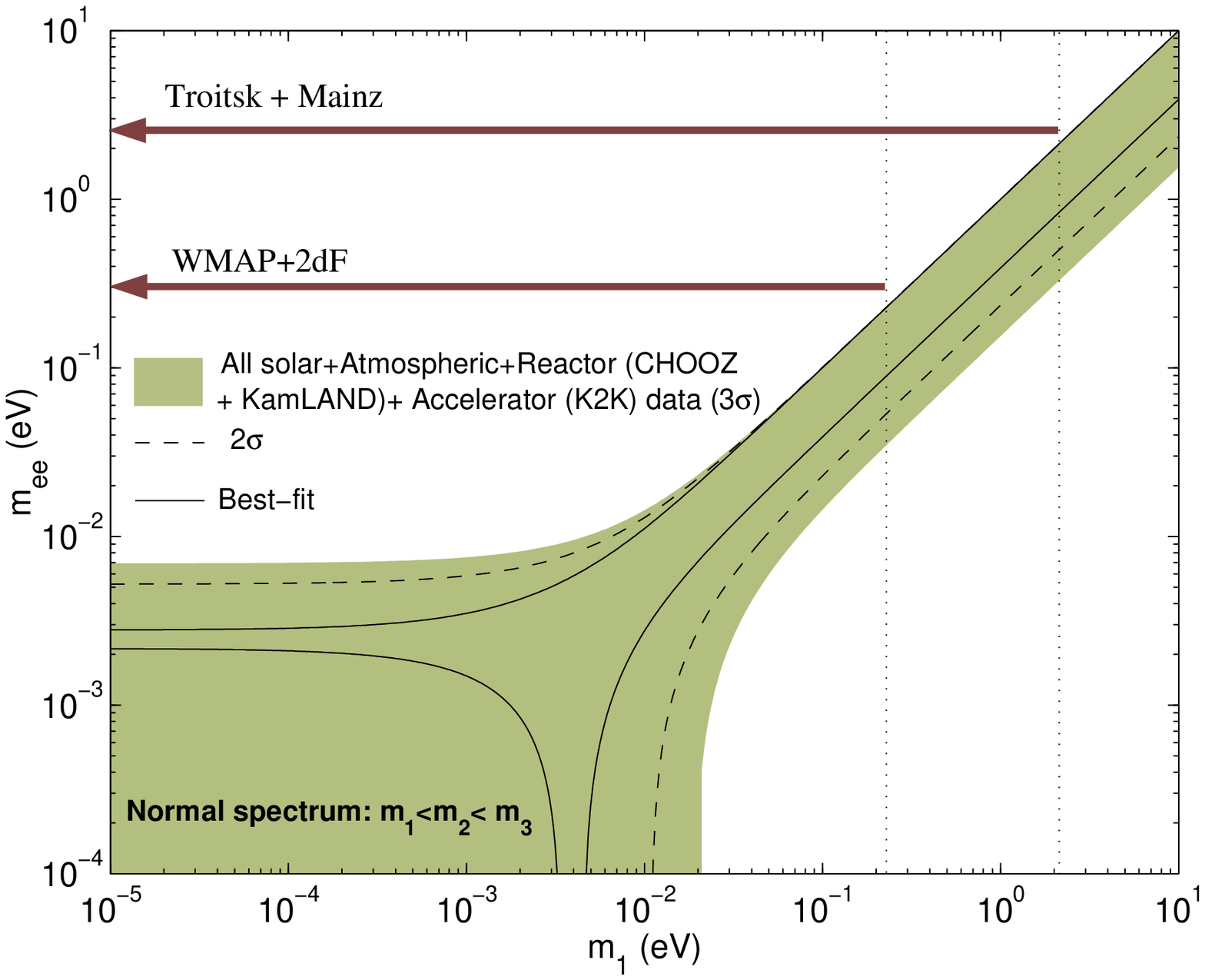}
&\includegraphics[width=8.3cm]{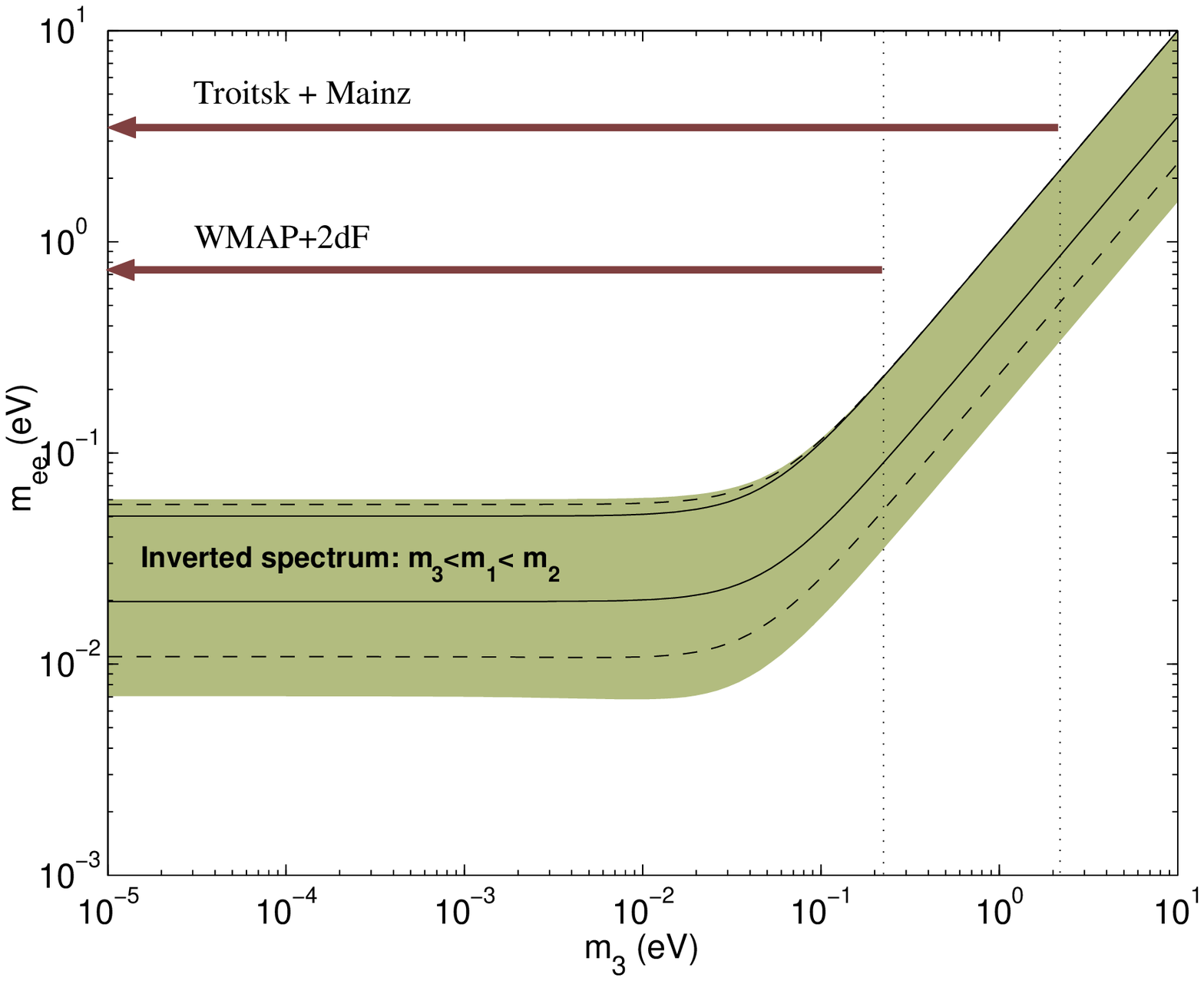}
\end{tabular}\label{fig1add}
\end{figure*}
In Fig.~\ref{fig1add} we show the updated plots for the allowed
ranges of $\mee$ as a function of the lightest neutrino mass,
using the $3\sigma$, $2\sigma$ and best-fit results given in
(\ref{valleresults}) and (\ref{bestsnovalle}). From the plots and
Eqs.~(\ref{boundHI})-(\ref{boundQD}), we get the following ranges
for $\mee$
\begin{align}
&  3\sigma\,(2\sigma)\,[\text{Best fit}]:\;\;\; 0\,(0)\,
[2.2]\times10^{-3} \lesssim \mee^{\rm HI} \lesssim 6.9\,(5.2)\,[2.8]\,\times10^{-3}\,\eV\,,\\
&  3\sigma\,(2\sigma)\,[\text{Best
fit}]:\;\;\;0.7\,(1.1)\,[2.0]\times10^{-2}\,\eV \lesssim \mee^{\rm
IH} \lesssim 6.1\,(5.7)\,[5.0]\times10^{-2}\,\eV\,,\\
&  3\sigma\,(2\sigma)\,[\text{Best fit}]:\;\;\;\mee^{\rm
QD}\gtrsim 0.035\,(0.05)\,[0.09]\;\; \text{for}\,\,m=0.23\,\eV\,.
\end{align}
One can therefore see that the new SNO results reduce the allowed
ranges for the effective Majorana mass parameter $\mee$.
Nevertheless, the main conclusions taken in the previous version
of this paper remain practically unchanged. Significant deviations
from these results still require further improvements on the
determination of the neutrino oscillation and $\ndbd$ parameters.

\bigskip \textbf{Acknowledgements}\\
The author thanks G.C. Branco and R. Gonz\'alez Felipe for the
careful reading of the manuscript and valuable suggestions. This
work was supported by {\em Funda\c{c}{\~a}o para a Ci{\^e}ncia e a Tecnologia}
(FCT, Portugal) under the grant \mbox{PRAXISXXI/BD/18219/98}.

\end{document}